\documentclass[final]{amsart}
\usepackage{amsmath,cite}
\def\C{\mathbb{C}}
\def\R{\mathbb{R}}
\def\Z{\mathbb{Z}}
\def\p{\mathbb{P}}
\def\id#1{\mathbb{I}_{#1}}

\def\rt{\longrightarrow}
\def\eq#1{(\ref{#1})}
\def\dual#1{{#1}^{\vee}}
\let\goth\mathfrak
\def\diag#1{\qopname\relax o{Diag\{{#1}\}}}
\def\ker#1{\qopname\relax o{Ker}~{#1}}
\def\tr#1{\qopname\relax o{Tr}~{#1}}
\def\coker#1{\qopname\relax o{coKer}~{#1}}
\def\hom#1{\qopname\relax o{Hom({#1})}}
\def\spec#1{\qopname\relax o{Spec}~{#1}}
\def\cf#1#2#3#4{\bibitem{#1}{#2}~{\it #3}\,;~{#4}.}

\def\FI{Fayet--Iliopoulos~}
\def\hep#1{hep-th/{#1}}
\def\alg#1{alg-geom/{#1}}
\def\npb#1#2#3{Nucl. Phys. {\bf B{#1}}~({#2})~{#3}}
\def\prd#1#2#3{Phys. Rev. {\bf D{#1}}~({#2})~{#3}}
\title{Conifolds from D-branes}
\author{Subir Mukhopadhyay} 
\address[Subir Mukhopadhyay]{Institute of Physics
\\Bhubaneswar 751 005, India }
\email{subir@iopb.stpbh.soft.net}
\author{ Koushik Ray}
\address[Koushik Ray]{INFN --- Sezione di Roma, `` Tor Vergata" \\ 
Via della Ricerca Scientifica, 1 00133 Roma, ITALY }
\email{koushik@roma2.infn.it}
\begin{document}
\begin{abstract}
In this note 
we study the resolution of conifold singularity by D-branes by considering 
compactification of D-branes on $\C^3/(\Z_2\times\Z_2)$. The resulting vacuum
moduli space of D-branes is a toric variety which turns out to be a resolved
conifold, that is a nodal variety in $\C^4$. This has the implication 
that all the corresponding phases of Type--II string theory are geometrical 
and are accessible to the D-branes, since they are related by flops. 
\end{abstract}
\date{\today}
\maketitle
In the aftermath of the {\em second superstring revolution} 
the role of D-branes in the study of space-time has been widely appreciated.
It has become quite 
apparent that one can probe sub-stringy scales in space-time using
D-branes --- the BPS states that carry Rammond-Rammond charges in
string theory. D-branes have recently been used to understand the
short-distance geometry of Calabi--Yau manifolds \cite{dm,jm,dgm}. 
These studies discern rather directly, as compared to the studies with 
fundamental strings, the phase structure of Calabi--Yau manifolds 
\cite{dm,dgm}. It has been shown that the topological and 
geometric properties of Calabi--Yau manifolds studied using both D-branes 
and fundamental strings corroborate.
More specifically, it has been
shown \cite{dgm} that the D0-branes do not  allow for
non-geometric phases to appear in the theory which is in keeping with the
conclusion drawn earlier 
from the analysis of M--theory and F--theory \cite{wit1}. 

The analysis of \cite{dgm} was for $\C^3/\Gamma$, where $\Gamma$
is a discrete Abelian group in $SU(3)$. Specific 
examples of $\Gamma=\Z_3$ and $\Gamma=\Z_5$ were worked out, which could
be generalized for any $\Z_n$. It was concluded that the D-branes probe
resolved $\C^3/\Z_n$ singularities. In this scheme one starts with
a set of D-branes on $\C^3$, arranged according to
the regular representation of $\Gamma$,
followed by a truncation to the $\Gamma$--invariant sector of the action,
thus ending up with a gauged sigma-model whose configuration
space, in the low-energy limit, is interpreted as the sub-stringy 
space-time. This picture has a
mathematical counterpart in terms of blowing up a singular K\"ahler
quotient \cite{inf1,inf2} in an orbifold, viewed as a toric variety. 

In this note we shall study another example of this construction, namely a
blown-up $\C^3/(\Z_2\times\Z_2)$, following the lines of \cite{dgm}. We shall
find that, in this case, the D-brane probe senses a double-point singularity,
which gets resolved for a generic value of the resolution parameter; 
and under a smooth variation of the resolution parameter one moves on
from one resolution to another, which corresponds to a flop \cite{can,gre}. 
\section{Moduli space of D-branes} 
The moduli space of Calabi--Yau compactification has been studied in 
different ways including either the construction of the mirror
manifold and thereby studying its complex structure moduli space 
\cite{agm}, or the 
construction of moduli space of the linear sigma model and subjecting it
to the action of a renormalization group in order to yield a
conformal-invariant physical space \cite{wit2,asgr}.
Both approaches yield a moduli space that has regions which either have 
a geometric interpretation or at least can be connected to a geometric phase
by means of an analytic continuation. Since D-branes are {\em natural} 
probes of (sub-stringy) space-time, it is important to study the geometric
structure sensed by them. 

This kind of study differs from the conventional
string theoretic studies in that, while a classical background geometry is
incorporated in an ad hoc fashion into string theory, and is then 
subjected to various corrections, D-brane geometry {\em emerges} 
from the study of the moduli space of the vacuum configuration
of D-branes \cite{jm}. Thus, at the regime of weak coupling, the 
geometry could be understood through a classical study
of the vacuum of the world-volume moduli space of the corresponding 
supersymmetric Yang-Mills theory. This world-volume theory turns out to be 
classically described by a gauged linear sigma model giving rise to 
space-times avoiding the occurrence of 
non-geometric phases. This moduli space is likely to admit quantum
correction in general, but for D0-branes at weak coupling those 
corrections are 
suppressed \cite{dgm}. Further, if we expect, following \cite[and also
\cite{dgm}]{bfss},
that there is no surprise lurking in the strong coupling regime, and 
extrapolate this result to strong coupling, 
then this is in conformity with the picture that arises in
considerations with M-theory \cite{wit1}. In this note we shall
discuss D-branes on an orbifold of $\C^3$ obtained by adding 
the moduli from the twisted sector of the theory with small coefficients. 

D$p$-branes near an orbifold singularity is represented by a super 
Yang-Mills Lagrangian with \FI term due to coupling 
to the closed string twisted sector \cite{dm}. It has been shown 
from world-sheet consistency conditions that the orbifold prescription
of the closed string theory also applies to the open string theory, with
new open string sectors provided by the images of the original
D-brane \cite{pol}. 
Thus a D$p$-brane
at a point in $\C^3/\Gamma$ will be defined as the quotient of the 
theory of $|\Gamma|$ number of D$p$-branes in $\C^3$ by a simultaneous 
action of $\Gamma$ on $\C^3$ and the D-brane index. Here $|\Gamma |$ 
denotes the dimension of the discrete group $\Gamma$. 

Thus, in order to construct the Lagrangian of D-branes compactified on 
an orbifold $\C^3/\Gamma$ of $\C^3$, where $\Gamma=\Z_2\times\Z_2$ is a 
third order discrete subgroup of $SU(3)$, we start with a 
theory of four D-branes in $\C^3$. In view of the observation that
the world-volume theory of D0-branes is the dimensionally reduced
ten-dimensional $N=1$ super-Yang-Mills theory, the present one
is equivalent to the dimensionally reduced $N=4$ super Yang-mills field 
theory in four dimensions with gauge group $U(n)$. The branes are
arranged in $\C^3$ in a $\Gamma$--symmetric fashion to begin with. 
Any generator of the discrete group $\Gamma$ acts on the 
space $\C^3$, as well as on the Chan-Paton degrees of freedom of 
the branes, the latter action being in the regular representation. 
We then quotient out the theory by $\Gamma$ by
restricting to the $\Gamma$--invariant sector of the theory. 

Let us begin by describing the action of $\Gamma$ on the bosonic fields.
The corresponding fermionic partners can be incorporated at any stage 
of the calculation cashing in on supersymmetry.
As there are four D-branes, they are represented by $4\times 4$ matrices
transforming in the adjoint of $U(4)$. For the sake of convenience 
we write the bosonic fields corresponding to the coordinates
tangential to the $\C^3$ as
a set of three $4\times 4$ complex matrices,
$\{X^1, X^2, X^3\}$. We shall then quotient them out by the discrete group
$\Gamma=\Z_2\times\Z_2$. Let us denote the three
generators of $\Z_2\times\Z_2$ by $\goth{g}_1$, $\goth{g}_2$ and 
$\goth{g}_1\goth{g}_2$. The action of the generators $\goth{g}_1$ and 
$\goth{g}_2$ on the $X$'s are given as: 
\begin{eqnarray}
X^\mu\rt R^\mu_\nu\gamma{(\goth{g}_i)}X^\nu\gamma{(\goth{g}_i)}^{-1},
\end{eqnarray}
where $R^{\mu}_{\nu}$ denotes a rotation matrix associated to the
generators, defined as
\begin{eqnarray}
R^\mu_\nu(\goth{g}_1) = \diag{-1,-1,+1}, \quad 
R^\mu_\nu(\goth{g}_2) = \diag{-1,+1,-1};  
\end{eqnarray}
while we have defined the matrices
\begin{eqnarray}
\gamma(\goth{g}_1) = \diag{1,-1,1,-1},
\quad\gamma(\goth{g}_2) = \diag{1,1,-1,-1}, 
\end{eqnarray}
with $\gamma$ denoting a regular representation of the 
corresponding generator written as its argument. For the fields 
corresponding to the coordinates transverse
to the $\C^3$ in question, the action is the same except that they are 
represented by real matrices and the rotation matrix is the identity: 
$R^\mu_\nu = \delta^\mu_\nu$. The fermionic fields are also
represented by $4\times 4$ matrices with a similar action of $\Gamma$ on them.

On quotienting  by $\Z_2\times\Z_2$ with the above-mentioned action,
the fields corresponding to the coordinates transverse to $\C^3$
become diagonal, while the three fields for the tangential ones assume the 
following forms:
\begin{eqnarray}\label{def:xyz}
X^1 = 
\begin{pmatrix}
0 & 0 & 0& x_1 \\ 0&0&x_2&0 \\0&x_3&0&0\\x_4&0&0&0
\end{pmatrix},
\quad X^2 = 
\begin{pmatrix}
0 & y_1 & 0& 0 \\ y_2&0&0&0 \\0&0&0&y_3\\0&0&y_4&0
\end{pmatrix},
\quad X^3 = 
\begin{pmatrix}
0 & 0 & z_1 & 0 \\ 0&0&0&z_2 \\z_3&0&0&0\\0&z_4&0&0
\end{pmatrix},
\end{eqnarray}
where $x_i$, $y_i$ and $z_i$, $i=1,2,3,4$, are complex numbers.
This leaves a subgroup $U(1)^4$ of $U(4)$ unbroken. An overall
$U(1)$ aside, we thus have a gauge group $U(1)^3$, which is further 
broken by the non-zero expectation values of the chiral fields, 
$X^1$, $X^2$ and $X^3$. Thus, the resulting theory represents a single 
D-brane in the orbifold, $\C^3/(\Z_2\times\Z_2)$.

In order to take into
account the coupling of the D-brane to the closed string twisted sector of 
the theory, one needs to introduce \FI terms in the Lagrangian. Since in 
this case there are three generators of $\Gamma$, we get three real
scalars from the NS-NS sector which are known to be responsible for 
blowing up the orbifold singularity \cite{dm,jm}. World-sheet 
considerations \cite{dm} indicate that in the presence of a twisted sector 
the theory picks up \FI terms such that the 
coefficient $\zeta_i$ of the $i$-th \FI term
corresponds to the $i$-th of the three $U(1)$'s and the value of each
$\zeta_i$ is determined by the expectation value of the NS-NS moduli fields.

As explained earlier, the space-time sensed by the D-brane can be derived 
from the moduli space of the above theory. Since we are
interested in the supersymmetric vacua of the theory, we impose the
D-flatness condition, namely,
\begin{eqnarray}\label{zeta1}
\zeta_1 &=& -|x_1|^2 + |x_4|^2 - |y_1|^2 + |y_2|^2 - |z_1|^2 + |z_3|^2, \\
\zeta_2 &=& -|x_2|^2 + |x_3|^2 - |y_2|^2 + |y_1|^2 - |z_2|^2 + |z_4|^2, 
\quad\mathrm{and}\\
\label{zeta3}
\zeta_3 &=& -|x_3|^2 + |x_2|^2 - |y_3|^2 + |y_4|^2 - |z_3|^2
+ |z_1|^2. 
\end{eqnarray}
Let us note that the coefficients $\{\zeta_i| i=1,2,3\}$ satisfy the condition 
$\sum_{i=1}^3 \zeta_i= 0$,
as is required for the existence of supersymmetric vacua \cite{dgm}. 

Let us now consider the superpotential of the theory, which is 
given by
\begin{eqnarray}
\mathcal{W} = \tr{[X^{1}, X^{2}] X^{3}}.
\end{eqnarray}
The vacuum satisfies the following condition,
\begin{eqnarray}
\partial\mathcal{W} = 0, \qquad\mathrm{or~~equivalently,} \nonumber \\ 
\label{potcnstr}[X^\mu ,X^\nu ] =0.
\end{eqnarray}
We first evaluate the superpotential constraints. Using the expressions
\eq{def:xyz} in \eq{potcnstr}, we derive twelve constraint relations
on the twelve variables. However, out of these twelve constraint relations 
six could be seen to be independent, which we choose as the following ones:
\begin{eqnarray}
\begin{matrix}
x_1x_4 = x_2x_3, & x_1y_4 = y_1x_2, \\
y_1y_2 = y_3y_4, & x_1z_4 = z_1x_3, \\ 
z_1z_3 = z_2z_4, & y_1z_2 = z_1y_3.
\label{conrel}
\end{matrix}
\end{eqnarray}
One can derive the other six equations that follow from the superpotential 
constraint using these six. Thus, we are left with six independent variables
from the components of the $X$'s. 
The D-brane moduli space is constructed out of these independent 
components.
The classical D-brane moduli space is obtained for a fixed set of values
of the \FI coefficients $\zeta_i$, by imposing the relations
\eq{zeta1}--\eq{zeta3} and the superpotential constraint \eq{conrel} and
further quotienting out by the residual $U(1)^3$ gauge group.
Let us now describe this quotient construction following \cite{dgm}. 
\section{Topological properties}
We want to find out the toric variety that describes the moduli space of
D-branes obtained via the quotient construction mentioned above. 
In other words, we 
want to describe the space of solutions of the superpotential constraints, 
after quotienting out by the residual $U(1)^3$ gauge group.

In the present case we are dealing with the  three-(complex)-dimensional 
moduli space of the D-branes. This can be described as a quotient space
($\C^9 -  \goth{F}_{\Delta})/(\C^{\star})^6$ for some set 
$\goth{F}_{\Delta}$ determined by the restricted set of values of $\zeta_i,
i=1,2,3$, which are relevant in this case and an action of $(\C^{\star})^6$
on the ambient $\C^9$.  Different choices of values of $\zeta_i, i=1,2,3$ 
correspond to different phases of the theory, which are related to the set 
$\goth{F}_{\Delta}$ to be removed from the ambient space $\C^9$. 

In order to construct a toric variety one needs a lattice and a fan 
\cite{fulton}, which is a set of strongly convex rational polyhedral
cones, $\{\goth{s}\}$. For the case at hand we take the lattice as
$\goth{M}=\Z^6$ and then choosing 
the independent variables as $\{x_1, y_1, y_2, z_1, z_2, z_4\}$,
we define the cone $\goth{s}$ as being generated by the 
the indices of the independent variables in the expressions of the twelve 
variables given in a matrix form as follows:
\begin{eqnarray}
\bordermatrix{
\nonumber
& x_1 & y_1 & z_1 & y_2 & z_2 & z_4 \\ 
\nonumber
x_1 & 1& 0 &0&0&0&0 \\
\nonumber
x_2 & 1&-1&1&1&-1&0 \\
\nonumber
x_3 & 1&0&-1&0&0&1 \\
\nonumber
x_4 & 1&-1&0&1&-1&1 \\
\nonumber
y_1 &0&1&0&0&0&0 \\
\nonumber
y_2 &0&0&0&1&0&0&\\
\nonumber
y_3 &0&1&-1&0&1&0\\
\nonumber
y_4 &0&0&1&1&-1&0\\
\nonumber
z_1 &0&0&1&0&0&0\\
\nonumber
z_2 &0&0&0&0&1&0\\
\nonumber
z_3 &0&0&-1&0&1&1\\
z_4 &0&0&0&0&0&1
}.
\end{eqnarray}
In other words, the cone $\goth{s}$ is generated by the rows of the 
rectangular $12\times 6$
matrix. Now to describe the toric variety, it is particularly 
convenient to consider the dual cone 
$\dual{\goth{s}}$. The dual cone is defined in a lattice 
$\dual{\goth{M}}=\hom{\goth{M},\Z}$, dual to $\goth{M}$. 
One writes the twelve variables 
$\{ x_i, y_i, z_i | i=1,2,3,4\}$ in terms of nine homogeneous 
coordinates $\{p_i | i=0, \cdots 8 \}$
\cite[also \cite{dgm}]{cox}. 
Indices of the homogeneous coordinates, as they appear in the monomials 
defining the original variables, determine the dual cone $\dual{\goth{s}}$. 
The columns of the matrix $T$ specifying a map 
\begin{eqnarray}
T:~ \C\,[p_0,p_1,p_2,p_3,p_4,p_5,p_6,p_7,p_8]\rt
\C\,[ x_1,y_1,y_2,z_1,z_2,z_4,
x_1^{-1},y_1^{-1},y_2^{-1},z_1^{-1},z_2^{-1},z_4^{-1}], 
\end{eqnarray}
generate the cone $\dual{\goth{s}}$:
\begin{eqnarray}
\label{mapt}
T=\bordermatrix{
\nonumber
&p_0 &p_1 &p_2 &p_3 &p_4 &p_5 &p_6 &p_7 &p_8 \\
\nonumber
x_1 & 1& 1& 1& 0& 0& 0& 0& 0& 0 \\
\nonumber
y_1 & 0&1&0&  1&1&0& 0&0&0 \\
\nonumber
y_2 & 0&0&0&  0&0&1& 1&1&0 \\
\nonumber
z_1& 1&0&0& 1&1&0& 0&0&0\\
\nonumber
z_2& 1&0&0& 0&0&0& 1&1&0\\
z_4& 0&0&0& 1&1&0& 0&0&1
}.
\end{eqnarray}
The transpose of the kernel of $T$, which is a $3\times 9$ matrix, 
given by
\begin{eqnarray}
\label{kert}
(\ker{T})^{\mathrm{T}}  = 
\begin{pmatrix}
-1 &-1&2& 0&1&-1& 0&1&-1 \\
-1 &-1&2& 1&0&-1& 1&0&-1 \\
-1 &-1&2& 1&0&-1& 0&1&-1 
\end{pmatrix},
\end{eqnarray}
then
specifies the action of the algebraic torus $(\C^{\star})^3$ on $\C^9$.
However, for our case we have to further take into account the fact that the
whole variety thus obtained is not relevant. The relevant part is what is
left over after quotienting by the $U(1)^3$ arising in the twisted 
sector of the moduli space. In order to dispense with the irrelevant part,
one has to consider the inclusion of an extra $(\C^{\star})^3$ in $\C^9$. 
As a whole, this is tantamount to prescribing an action of
$(\C^{\star})^6$ on the toric variety. Since the set 
$\C^9 - \goth{F}_{\Delta}$ 
contains $(\C^{\star})^9$, the toric variety $(\C^9 - \goth{F}_{\Delta}) / 
(\C^{\star})^3$ contains $(\C^{\star})^9/(\C^{\star})^3 \simeq (\C^{\star})^6$ 
as a  dense open subset. As a group this acts on the toric variety through an 
action of $(\C^{\star})^6$ on $\C^9$, provided the projection of  the action 
to $(\C^{\star})^9/(\C^{\star})^3 \simeq (\C^{\star})^6$ gives the identity map.
Such an action is specified by a $6\times 9$ charge matrix, denoted $U$, 
and the condition on the projection is, 
\begin{eqnarray}
TU^{\mathrm{T}} = \id{6},
\end{eqnarray}
where $T$ is the $6\times 9$ matrix defined in \eq{mapt}. 
In order to find out the matrix $U$, we thus have to invert the rectangular 
matrix $T$. We use the Moore-Penrose prescription which gives a unique inverse.
The matrix $U^{\mathrm{T}}$ is evaluated as the transpose of the 
Moore-Penrose inverse of the transpose of $T$. It takes the following form:
\begin{eqnarray}
U=\begin{pmatrix}
   {2/ 9} & {2/ 9} & {5/ 9} & -{1/ 9} & -{1/ 9} & 
   {2/ 9} & -{1/ 9} & -{1/ 9} & {2/ 9} \cr -{1/ 3} & 
   {2/ 3} & -{1/ 3} & {1/ 6} & {1/ 6} & -{1/ 3} & 
   {1/ 6} & {1/ 6} & -{1/ 3} \cr -{1/ 9} & -{1/ 9} & 
   {2/ 9} & {1/ {18}} & {1/ {18}} & {8/ 9} & {1/ {18}} & 
   {1/ {18}} & -{1/ 9} \cr {1/ 2} & -{1/ 2} & 0 & {1/ 4}
    & {1/ 4} & {1/ 2} & -{1/ 4} & -{1/ 4} & -{1/ 2} \cr 
   {1/ 6} & {1/ 6} & -{1/ 3} & -{1/ {12}} & -{1/ {12}} & 
   -{5/ 6} & {5/ {12}} & {5/ {12}} & {1/ 6} \cr -{1/ 9} & 
   -{1/ 9} & {2/ 9} & {1/ {18}} & {1/ {18}} & -{1/ 9} & 
   {1/ {18}} & {1/ {18}} & {8/ 9} \cr  
\end{pmatrix} .
\end{eqnarray}
In order to describe the (symplectic or holomorphic) quotient of $\C^9$ with 
respect to $U(1)^3$, that is $\C^9//U(1)^3$, we need specify the charges
through which the independent fields couple to the three $U(1)$ gauge fields
associated with the three
\FI terms. These are given by 
\begin{eqnarray}
V = \begin{pmatrix}
-1 & -1&1&-1&0&0 \\
0&1&-1&0&-1&1\\
0&0&0&1&0&0
\end{pmatrix}.
\end{eqnarray}
Then multiplying $V$ and $U$ one derives,
\begin{eqnarray}
VU = 
\begin{pmatrix}
 -{1/ 2} & -{1/ 2} & 0 & -{1/ 4} & -{1/ 4} & {1/ 2} & {1/ 4} & {1/ 4} & {1/ 2} \\ -{1/ 2} & {1/ 2} & 0 & {1/ 4} & {1/ 4} & -{1/ 2} & -{1/ 4} & -{1/ 4} & {1/ 2} \\ {1/ 2} & -{1/ 2} & 0 & {1/ 4} & {1/ 4} & {1/ 2} & -{1/ 4} & -{1/ 4} & -{1/ 2} \\  
\end{pmatrix}.
\end{eqnarray}
Multiplying $VU$ with an innocuous factor of 4 and then concatenating
with the charge matrix \eq{kert} one gets, after 
some row-operations, the following matrix describing the embedding 
of the algebraic torus $(\C^\star)^6$ in $\C^9$. 
\begin{eqnarray}
\label{kiu}
\widetilde{Q} = \begin{pmatrix}
-1 &-1 &2 &0 &1 &-1 &0 &1 &-1\\
-1 &-1 &2 &1 &0 &-1 &1 &0 &-1\\
-1 &-1 &2 &1 &0 &-1 &0 &1 &-1\\
 -2 & -2 & 0 & -1 & -1 & 2 & 1 & 1 & 2 \\ -2 & 2 & 0 & 1 & 1 & -2 & -1 & -1 & 2 \\ 2 & -2 & 0 & 1 & 1 & 2 & -1 & -1 & -2 \\  
\end{pmatrix}  .
\end{eqnarray}
The co-kernel of the transpose of $\widetilde{Q}$, after ignoring the 
repetitions, takes the following form:
\begin{eqnarray}
\widetilde{T} = \coker{\widetilde{Q}^{\mathrm{T}}} = 
\begin{pmatrix}
0& 0&1&-1\\
0&1&0&1\\
1&0&0&1
\end{pmatrix};
\end{eqnarray}
and further, one can obtain the kernel of $\widetilde{T}$ to be:
\begin{eqnarray}
\label{kerttil}
\ker{\widetilde{T}} = \begin{pmatrix}
-1 \\ -1\\ 1\\ 1
\end{pmatrix} .
\end{eqnarray}
Using  \eq{kerttil}
one can define the semigroup $\goth{S}_{\goth{s}}$ 
generated finitely by the toric ideal $\langle xy-zw\rangle$ 
\cite{fulton}.
The corresponding ``group" ring  can be written as
\begin{eqnarray}
\C\,[\goth{S}_{\goth{s}}] = \C\,[x,y,z,w] / \langle xy-zw\rangle;
\end{eqnarray}
and therefore one derives, 
\begin{eqnarray}
\spec{\C\,[\goth{S}_{\goth{s}}]} = xy - zw,
\end{eqnarray}
 where $x,y,z,w$ are four complex numbers.
Recalling that the equation of a conifold in $\C^4$  can be cast in the 
form \cite[for example]{ossa}
\begin{equation}
\label{coni}
xy - zw = 0, \qquad\mathrm{with}~~x,y,z,w\in \C,
\end{equation}
one concludes that the toric variety defined  by  $\widetilde{T}$ is 
the blow-up of the conifold \eq{coni}. Blow-up of the conifold \eq{coni} has 
been studied in detail in \cite{ossa}. 

The above moduli space can also be realized in terms of the 
alternative description of the 
gauged linear sigma model due to Witten. The homogeneous variables 
$\{p_i\}$ play the role of the chiral superfields which are charged with 
respect to the gauge group $U(1)^3$. The charge of the $i$-th field with 
respect to  the $j$-th $U(1)$ is given by the entry $\widetilde Q_{ij}$
of the charge matrix \eq{kiu}. The \FI parameters are again given by 
$\zeta_i$ associated only with the $U(1)$'s coming from the original 
world volume theory. The coefficients corresponding to 
the other $U(1)$'s vanish. The moduli space again follows from the D-term
equation corresponding to this model. This is known as symplectic quotienting
\cite{wit2,asgr}, where one views $\C^\star$ as $\R_+\times U(1)$. Apart from 
imposing the D-term equation, one has to again mod out the 
remaining three $U(1)$'s to get the moduli space, as before.
 
In order to describe the moduli space in a simple form, let us start 
with the charge matrix $\widetilde Q$ with the \FI parameters adjoined 
to it. So the matrix becomes 
\begin{eqnarray}
\widetilde{Q} = \begin{pmatrix}
-1 &-1 &2 &0 &1 &-1 &0 &1 &-1 &0\\
-1 &-1 &2 &1 &0 &-1 &1 &0 &-1 &0\\
-1 &-1 &2 &1 &0 &-1 &0 &1 &-1 &0\\
 -2 & -2 & 0 & -1 & -1 & 2 & 1 & 1 & 2 & \zeta_1\\ 
-2 & 2 & 0 & 1 & 1 & -2 & -1 & -1 & 2 & \zeta_2\\ 
2 & -2 & 0 & 1 & 1 & 2 & -1 & -1 & -2 & \zeta_3\\
\end{pmatrix}.
\end{eqnarray}
After some row-operations we obtain,
\begin{eqnarray}
\widetilde{Q} = \begin{pmatrix}
0 &0 &0 &0 &-2 &0 &0 &+2 &0 &-\frac{\zeta_2+\zeta_3}{2}\\
0 &0 &0 &0 &0 &0 &1 &-1 &0 &0\\
0 &0 &0 &1 &-1 &0 &0 &0 &0 &\frac{\zeta_2+\zeta_3}{4}\\
 0 & 0 & -4 & 0 & 0 & 4 & 0 & -4 & 4 & \zeta_1+\zeta_2+\zeta_3\\ 
-4 & 0 & 0 & 0 & 0 & 0 & 0 & 0 & 4 & \zeta_1+\zeta_2\\ 
0 & -4 & 0 & 0 & 0 & 4 & 0 & 0 & 0 & \zeta_1+\zeta_3\\  
\end{pmatrix}.
\end{eqnarray}
The equations ensuing from all the rows of the matrix $\widetilde{Q}$, 
except the fourth one, merely express 
one variable in terms of the others. Hence these  do not impose any 
non-trivial 
constraint on the resulting space. One can just trade one variable for 
the other and thus, finally, the independent variables to consider
reduces to $p_2$, $p_5$, 
$p_7$ and $p_8$. They are constrained by the non-trivial equation given by 
the fourth row which is, 
\begin{equation}
|p_2|^2 - |p_5|^2 + |p_7|^2 - |p_8|^2 
= - \frac{(\zeta_1 +\zeta_2 +\zeta_3)}{4}.
\end{equation}
When all the moduli parameters vanish, this becomes a conifold with 
a node, while for any generic value the node gets resolved into a 
${\mathcal O}(-1)\times{\mathcal O}(-1)$-bundle over 
the one-dimensional projective space $\p^1$. So crossing the 
singularity corresponds to a flop transition.
\section{Discussion}
In this note we have considered certain topological properties of D-brane 
moduli space near an Abelian orbifold singularity, $\C^3/(\Z_2\times\Z_2)$. 
We have found that the moduli describe a space with a resolved double-point 
singularity. The \FI parameters originating in the
closed string twisted sector control the size of the blow-up. 

In this analysis we have taken the metric to be flat. We certainly wish 
that the final metric on the blown-up conifold were Ricci-flat and matches 
the one that can be obtained from low-energy classical supergravity 
considerations, valid for small values of the curvature, {\em i.e.}
$R << 1/l_p^2$, where $l_p$ denotes the Planck length. However, as 
is the case with $\Z_3$ or $\Z_5$ \cite{dgm},
it might happen in certain cases that the metric on the resolution is not 
Ricci flat: perhaps only the bound states of a large number of D-branes will
sense a Ricci flat metric \cite{dgm}. Therefore, it would be  interesting 
to investigate whether in the present case the metric on the resolved conifold
is Ricci flat or not. Such a metric has been found earlier in studying 
conifolds \cite{ossa}. Work in this direction is in progress.

In the present example we have got the resolutions which are connected by 
flops. A similar construction for the drastic conifold transition with 
D-branes may help understanding the topology changing process from the 
short-distance viewpoint.

It would also be interesting to generalize these considerations for the $D_n$ 
and $E_n$ series of groups. At the moment it is not clear whether for these 
non-Abelian groups one can even visualize the moduli space as a toric variety. 
If $\Gamma$ does not belong to $SU(3)$, then the resulting theory will be a 
non-supersymmetric theory and it may be interesting
to investigate those theories too. 

\vspace{0.5cm}
\noindent {\bf Note Added}: After completion of this work two related
papers appeared \cite{mut,brg}. One of these \cite{brg} has overlap 
with ours.

\vspace{0.5cm}
\noindent {\bf Acknowledgement}: It is a pleasure to thank A. Kumar 
for stimulating discussions and useful suggestions.

\end{document}